\documentclass[11pt,letterpaper]{article}
\usepackage{setspace}
\usepackage{color}
\usepackage[dvipsnames]{xcolor}
\usepackage{amsmath}
\usepackage{amsfonts}
\usepackage{bbm}
\usepackage{subcaption}
\usepackage{comment}
\usepackage{booktabs}
\usepackage{verbatim}
\usepackage{amssymb}
\usepackage{physics}
\usepackage{graphicx}
\usepackage{hyperref}
\usepackage[toc,page]{appendix}

\numberwithin{equation}{section} 
\providecommand{\U}[1]{\protect\rule{.1in}{.1in}}
\begin{document}
\title{\textbf{Skyrmions at Finite Density}}
\author{Fabrizio Canfora$^{1,2}$, Scarlett C. Rebolledo-Caceres$^{3}$ \\
{\small $^{1}$\textit{Facultad de Ingeniería y Tecnología,
Universidad San Sebastían, General Lagos 1163, Valdivia
5110693, Chile,}}\\
{\small $^{2}$\textit{Centro de Estudios Cient\'{\i}ficos (CECS), Casilla
1469, Valdivia, Chile,}}\\
{\small $^{3}$\textit{Departamento de F\'{\i}sica, Universidad de Concepci%
\'{o}n, Casilla 160-C, Concepci\'{o}n, Chile.}}\\
{\footnotesize fabrizio.canfora@uss.cl \& srebolledo2017@udec.cl}}
\maketitle
\begin{abstract}
In this paper, we will describe recent advances in analytical methods to
construct exact solutions of the Skyrme model (and its generalizations)
representing inhomogeneous Hadronic condensates living at finite Baryon
density. Such novel analytical tools are based on the idea to generalize the
well known spherical hedgehog ansatz to situations (relevant for the
analysis of finite density effects) in which there is no spherical symmetry
anymore. Besides the intrinsic mathematical interest to find exact solutions
with non-vanishing Baryonic charge confined to a finite volume, this
framework opens the possibility to compute important physical quantities
which would be difficult to compute otherwise.

\vspace*{\fill} 
\end{abstract}

\tableofcontents


\section{Introduction}

One of the main open problems in physics is the lack of an analytic
understanding of the low energy phase diagram of Quantum Chromodynamics (QCD
henceforth). Due to the non-perturbative nature of low energy QCD, many
reseacrhers believe that, in this regime, only refined numerical techniques
are useful (see \cite{newd3,newd4,newd5,newd6}, and references therein)
while analytical tools are not useful. One of the many negative consequences
of this fact is that, the appearance of the \textit{inhomogeneous Baryonic
condensates} (see \cite%
{pasta1,pasta2,pasta2a,pasta2b,pasta3,pasta4,pasta5,pastacond4,pasta7,pasta8,pasta9}%
, and the nice up to date review \cite{pasta10}), which is a very remarkable
phenomenon typical of finite Baryon density and low-temperature regime, has
no theoretical \textquotedblleft first principles" explanation. In such a
phase, ordered structures appear, in which most of the Baryonic charge is
contained in regular shapes like thick Baryonic layers or thick Baryonic
tubes. This phase has many similarities with the non-homogeneous
condensates, which have been discovered in integrable field theories in $%
(1+1)$ dimensions (see \cite{InCond1,InCond2,InCond3,InCond4,InCond}, and
references therein).

The numerical analysis of these configurations is quite demanding from the
hardware viewpoint (see \cite%
{aprox0,aprox1,aprox2,aprox3,aprox4,aprox5,aprox6,aprox7,aprox8,aprox9,aprox10}%
, and references therein). This unfortunate circumstance makes the study of
relevant physical quantities difficult. Transport coefficients (such as
thermal conductivity, viscosity, and so on) which encode fundamental
properties of multi-Baryonic configurations (see \cite%
{pastacond1,pastacond2,pastacond3,pastacond4,pastacond5}, and references
therein) are especially challenging. It is very complicated to analyze
numerically how these coefficients depend on the Baryonic charge, on the
coupling constants of the theory and so on. The present paper reviews a
novel approach able to describe analytically these inhomogeneous Baryonic
condensates (together with their corresponding electromagnetic fields)
appearing in the low energy limit of QCD opening the possibility to
determine exactly relevant physical quantities associated to these
configurations.

The starting point is the Skyrme theory which (at leading order in the t'
Hooft expansion \cite{Gerard,largeN1}, and \cite{largeN2}) represents the
low energy limit of QCD. The dynamical field of the Skyrme action \cite%
{skyrme} is a $SU(N)$-valued scalar field $U$ (here we will consider the
two-flavors case $U(x)\in SU(2)$ although many of the present results can be
extended to the generic $SU(N)$ case). This action possesses both small
excitations describing Pions and topological solitons describing Baryons 
\cite{Lizzi,shifman1,shifman2,witten0,ANW,Guadagnini}, and \cite{callan};
the Baryonic charge being a topological invariant (see also \cite%
{[8]p,[9]p,[10]p,[11]p,[12]p,sakurai,Machleidt,m1,m2,m3}, and references
therein).

At first glance, when one tries to analytically describe topological
solitons with high topological (Baryonic, in the present case) charge in
non-integrable theories (such as the low energy limit of QCD) a huge
technical problem appears. Generically, it is not possible to require
spherical symmetry for the SU(2)-valued field. In the present review, for
the reasons mentioned above, we are interested in finite density effects.
This, roughly speaking, means that we need to force the topological solitons
to live within a box of finite spatial volume (in which case spherical
symmetry is broken). In the usual textbook cases, the requirement of
spherical symmetry (in the hedgehog sense) generates an ansatz that both reduces the Skyrme field equations to just one equation as well as produces
a non-vanishing Baryon density. In fact, in the Skyrme case, spherical
symmetry is not so welcome\footnote{See \cite{Flambaum:2023yua,Leask1,Leask2} for an updated discussion on this point.} (from the analytic viewpoint) after all since the
remaining field equation can only be solved numerically. The situation
becomes even worse if the coupling with the electromagnetic field is taken
into account. The obvious reason is that two of the three Pions are charged
so that they generate their own electromagnetic field and, at the same time,
the electromagnetic field back-reacts on the $SU(2)$ field. In mathematical
terms, when the minimal coupling to the $U(1)$ gauge field is taken into
account, one has to solve a coupled system of 7 non-linear partial
differential equations (3 from the gauged Skyrme model and, in principle, 4
from Maxwell equations with the $U(1)$ current arising from the Skyrme model
itself). Consequently, the analytic construction of gauged inhomogeneous
condensates in the gauged Skyrme model Maxwell system looks as a hopeless
task.

Quite remarkably, as discussed in \cite%
{crystal1,crystal2,crystal3,crystal3.1,crystal4,crystal4.1,firstube,firsttube2,gaugsksu(n),modulation,56,56b,56a0,56a,56a1,56a11,56a2,56b1,56c}
the absence of spherical symmetry opens unprecedented possibilities as far
as analytic solutions are concerned. Such an approach provides (at least)
two explicit cases: Baryonic layers and tubes (together with their
modulations and their electromagnetic fields). It is worth emphasizing that,
first of all, the plots in \cite{crystal1} and \cite{modulation} are
qualitatively very close to the ones found numerically in the analysis of
spaghetti-like configurations (see the plots in \cite%
{pasta1,pasta2,pasta2a,pasta2b,pasta3}, and \cite{pasta10}). Moreover, in 
\cite{gaugsksu(n)} the shear modulus of lasagna configurations has been
computed, the result being in good agreement with \cite{pasta5} and \cite%
{pasta9}. A welcome consequence of the above formalism is that it opens the
possibility of analyzing the transport properties of these Hadronic tubes
and layers explicitly (as we will discuss at the end of the review).

The review is divided as follows: In section $2$, we give a brief
description of the gauged Skyrme Model, its relation with baryons, its field
equations, the energy-momentum tensor, and we define topological charge in
the model. Furthermore, we define and explain the most common ansatzes for the fundamental element $U$.

Section $3$ summarizes previous contributions of hadronic configurations at
finite density, we called these structures Skyrmions crystals. Then, we will present the metrics and their ranges of coordinates.

Section $4$ presents a novel analytic solution of the Skyrme Model at finite density with a non-vanishing topological charge. Besides, we define the relevant properties of the theory, like the energy-momentum tensor and the topological current, and present an interpretation of the chiral degrees of freedom.

Section $5$ presents the gauged version of the inhomogeneous condensates.

In section $6$, we show recent work related to a sector of pure Yang-Mills and Yang-Mills-Higgs theory that exhibits conformal symmetry. We motivate this work with the exciting possibilities of computing and characterizing nuclear matter.

Finally, section $7$ is dedicated to conclusions and future outlooks.

In our convention $c=\hbar =1$, Greek indices run over the space-time with mostly plus signature, and Latin indices are reserved for those of the internal space. For simplicity, we denote the scalar product with $\cdot$, for example, we use $(\nabla \alpha \cdot \nabla G)$ for $(\nabla_{\mu} \alpha \nabla^{\mu} G)$.



\section{The Skyrme-Maxwell Model}

A proper understanding of how topological solitons react to the presence of
non-trivial boundary conditions (such as the ones appearing then the theory
is analyzed in a box of finite spatial volume) is the key step to disclose
the mechanism behind the appearance of non-homogeneous Baryonic condensates.
A fundamental ingredient is that (at leading order in the 't Hooft expansion 
\cite{Gerard,largeN1},and \cite{largeN2}) the low energy limit of QCD is
equivalent to the gauged Skyrme model \cite{skyrme,callan,skyrme1}, and \cite{99}. An intriguing characteristic of Skyrme theory is that it
encodes both (topologically trivial) small excitations describing mesons
(Pions in the $SU(2)$ case) and topological solitons describing Baryons \cite%
{aprox3,aprox4,Lizzi,witten0,ANW,Guadagnini,callan,manton,BalaBook}, and \cite{99}. Moreover,
the Baryonic charge can be expressed as a topological invariant.
Skyrme theory has been analyzed at finite Baryon density starting from the
classic references \cite{aprox1,SkCr2,SkCr3,SkCr4}, and \cite%
{SkCr5} using either a variational approach or the so-called rational map
approximation. The techniques introduced in \cite{crystal1,crystal3,crystal4,gaugsksu(n),56a0,56a,56a1,56a2,56b1}, and \cite{56c} (which will be described in the
following sections) provided the first analytic constructions of
inhomogeneous Baryonic condensates in the low energy limit of QCD.

Needless to say, the gauged Skyrme model is a very important theory in
itself, but the 't Hooft large $N_{c}$ expansion gives rise to very
complicated subleading corrections to the Skyrme model (see \cite%
{subleading1,subleading2}, and \cite{subleading3} and references therein).
Therefore, an important question to address is:

Do such subleading terms ruin the approach developed in \cite{crystal1,crystal3,crystal4,gaugsksu(n),56a0,56a,56a1,56a2,56b1}, and \cite{56c}?

This issue has been discussed in \cite{crystal3,modulation}, and \cite%
{crystal5} where it has been shown that such an approach works, and it is almost
unchanged even when these subleading terms are included. In the present
review we will only explicitly consider the $SU(2)$ Skyrme model minimally
coupled to Maxwell theory.

The action for the gauged $SU(2)$-Skyrme model is given by 
\begin{equation}
I[U,A_{\mu }]=\int d^{4}v\left( \mathcal{L}^{\text{SK}}+\mathcal{L}^{\text{%
U(1)}}\right) \ ,  \label{actionskyrme1}
\end{equation}%
\begin{eqnarray*}
\mathcal{L}^{\text{SK}} &=&\frac{K}{4}\text{Tr}\left\{ R_{\mu }R^{\mu }+%
\frac{\lambda }{8}G_{\mu \nu }G^{\mu \nu }\right\} \ ,\ \mathcal{L}%
^{U(1)}=F_{\mu \nu }F^{\mu \nu }\ , \\
F_{\mu \nu } &=&\partial _{\mu }A_{\nu }-\partial _{\nu }A_{\mu }\ ,
\end{eqnarray*}%
where 
\begin{equation*}
R_{\mu }=U^{-1}D_{\mu }U=R_{\mu }^{a}t_{a}\ ,\quad G_{\mu \nu }=[R_{\mu
},R_{\nu }]\ ,\ d^{4}v=\sqrt{-g}d^{4}x\ ,
\end{equation*}%
\begin{equation*}
D_{\mu }U=\nabla _{\mu }U+A_{\mu }U\hat{O}\ ,\quad \hat{O}=U^{-1}\left[
t_{3},U\right] \ ,  \label{covdev}
\end{equation*}%
$U(x)\in SU(2)$, $g$ is the metric determinant, $\nabla _{\mu }$ is the
partial derivative, $D_{\mu }$ is the covariant derivative associated to the 
$U(1)$ gauge field $A_{\mu }$ and $t_{a}=i\sigma _{a}$ are the generators of
the $SU(2)$ Lie group, being $\sigma _{a}$ the Pauli matrices. The Skyrme
couplings $K$ and $\lambda $ are positive constants that have to be fixed
experimentally (see \cite{ANW,Guadagnini} and references therein).
The first term in Eq. \eqref{actionskyrme1} is the non-linear sigma model.
Skyrme introduced the second term to avoid Derrick's scaling argument \cite%
{Derrick} (which prevented the appearance of solitonic configurations in the
Non-Linear Sigma Model, NLSM henceforth). It is fair to emphasize that
actually, Skyrme understood the essence of Derrick's argument before Derrick
himself \cite{skyrme}.

\subsection{Field Equation and energy-momentum tensor}

The field equations for the Skyrme Model are obtained through its variation
with respect to the fundamental fields $U$ and $A_{\mu }$%
: 
\begin{eqnarray}
\nabla _{\mu }\left( R^{\mu }+\frac{\lambda }{4}[R_{\nu },G^{\mu \nu
}]\right) &=&0\ ,  \label{Eqoskyrme} \\
\nabla _{\mu }F^{\mu \nu } &=&J^{\nu },  \label{Eqomaxwell}
\end{eqnarray}%
where the electromagnetic current $J_{\mu }$ is given by 
\begin{equation*}
J_{\mu }=\frac{K}{2}\text{Tr}\left\{ \hat{O}\left( R_{\mu }+\frac{\lambda }{4%
}[R^{\nu },G_{\mu \nu }]\right) \right\} \ .
\end{equation*}

As has been already emphasized in the introduction, the choice of an
effective Ansatz is the most difficult step along the path to finding analytic
(gauged) Skyrmions due to the highly non-linear character of the field
equations.

The contribution of the Skyrme model to the energy-momentum tensor is given
by 
\begin{equation}
T_{\mu \nu }^{\text{SK}}=-\frac{K}{2}\text{Tr}\left\{ R_{\mu }R_{\nu }-\frac{%
1}{2}g_{\mu \nu }R^{\alpha }R_{\alpha }+\frac{\lambda }{4}\left( g^{\alpha
\beta }G_{\mu \alpha }G_{\nu \beta }-\frac{1}{4}g_{\mu \nu }G_{\alpha \beta
}G^{\alpha \beta }\right) \right\} \ ,  \label{TmunuSK}
\end{equation}%
while the contribution of the Maxwell theory is 
\begin{equation*}
T_{\mu \nu }^{U_{\left( 1\right) }}=g^{\alpha \beta }F_{\mu \alpha }F_{\nu
\beta }-\frac{1}{4}g_{\mu \nu }F^{\alpha \beta }F_{\alpha \beta }\ .
\label{TmunuMax}
\end{equation*}%
In most of the present review, we will focus on the energy-momentum tensor
of Skyrme Model $T_{\mu \nu }^{\text{SK}}$ because across this work we will
use $A_{\mu }\rightarrow 0$. On the other hand, we will mention how the
inhomogeneous Baryonic condensates described in the following sections can
be generalized with the inclusion of the minimal coupling with Maxwell
theory in due time.

\subsection{Topological Charge like a Baryon Number}

\label{ch1:sec:topologicalQ} Unlike Noether charges (which are continuous
conserved charges associated to symmetries and which are conserved
on-shell), topological charges are discrete, not associated to
symmetries and invariant under any continuous deformation of the fields
(so that these charges are conserved off-shell). In the case of gauged
Skyrme theory, the topological charge (which is interpreted as the Baryonic
charge of the given configuration) reads 
\begin{equation}
B=\frac{1}{24\pi ^{2}}\int_{\sigma }\rho _{B},  \label{Bcurrent}
\end{equation}%
where $\sigma $ is any three-dimensional $t=\text{const.}$ hypersurface, while the
Baryon density $\rho _{B}$ is defined as 
\begin{equation*}
\rho _{B}=\rho ^{\text{SK}}+\rho ^{\text{U(1)}} \, ,
\end{equation*}%
where 
\begin{align*}
\rho ^{\text{SK}}& =\epsilon ^{abc}\text{Tr}\left\{ \left( U^{-1}\partial
_{a}U\right) \left( U^{-1}\partial _{b}U\right) \left( U^{-1}\partial
_{c}U\right) \right\} , \\ 
\rho ^{\text{U(1)}}& =-\text{Tr}\left\{ \partial _{a}\Bigr(3A_{b}t_{3}\bigr(%
U^{-1}\partial _{c}U+\left( \partial _{c}U\right) U^{-1}\bigr)\Bigr)\right\} \, .
\end{align*}%
The topological charge is uniquely linked to the boundary conditions of the
fundamental fields $U$ and $A_{\mu }$. One of the most important properties
of this kind of charge is the fact that infinite energy is needed to change
a physical state with a fixed value of the topological charge to a state
with a different topological charge.

Whenever $A_{\mu }\rightarrow 0$ we get the topological charge associated to
the Skyrme model by itself. As it is well known since the introduction of
the model (see \cite{skyrme,skyrme1}, for a review see \cite{manton})
the energy of static configurations is bounded from below by the topological
charge:

\begin{equation*}
E\geq |Q| \, .
\end{equation*}%
Unlike what happens in Yang-Mills-Higgs theory, such a bound cannot be
saturated on flat space-time. This fact prevents one from using the common
techniques (associated to integrability and self-duality) to find analytic
solutions of the Skyrme field equations.

In what follows, we will use the terminology ``Skyrmions at finite
density" and ``inhomogeneous Baryonic condensates" to denote analytic
solutions of the field equations in Eqs. (\ref{Eqoskyrme}) and (\ref%
{Eqomaxwell})\ with non-vanishing topological charge satisfying boundary
conditions corresponding to a finite spatial volume.

\subsection{Ansatz for Skyrme Field}

In this section, we will show two explicit parametrizations of the $SU(2)$-valued Skyrme field $U$ which are very convenient to construct the sought ansatze.


\subsubsection{Exponential parametrization}

\label{sec:skygenericsphSU(2)}

The standard exponential form of the field $U(x)$ is 
\begin{equation}
U=\cos {(\alpha )}\mathbf{1}_{2\times 2}+\sin {(\alpha )}n_{j}t^{j},
\label{U1}
\end{equation}%
where $\mathbbm{1}_{2\times 2}$ is the $2\times 2$ identity matrix and $%
t^{i}=i\sigma ^{i}$ where $\sigma ^{j}$ are the Pauli matrices. Besides, 
\begin{equation}
\vec{n}=\left( \sin (\Theta )\cos (\Phi ),\sin (\Theta )\sin (\Phi ),\cos
(\Theta )\right) \, ,  \label{U2}
\end{equation}%
where $\alpha =\alpha (x^{\mu })$, $\Theta =\Theta (x^{\mu })$ and $\Phi
=\Phi (x^{\mu })$ are the three scalar degrees of freedom of the $SU(2)$%
-valued field $U(x)$.

Then, plugging Eqs. \eqref{U1} and \eqref{U2} into the action for the Skyrme
model reads 
\begin{align*}
I(\alpha ,\Theta ,\Phi )=\frac{K}{4}\int d^{4}v\text{Tr}\biggl \{& (\nabla
\alpha )^{2}+\sin ^{2}(\alpha )((\nabla \Theta )^{2}+\sin ^{2}(\Theta
)(D\Phi )^{2})+\frac{\lambda }{2}\bigr(\sin ^{2}(\alpha )((\nabla \alpha
)^{2}(\nabla \Theta )^{2} - \notag \\
& -(\nabla \alpha \cdot \nabla \Theta )^{2})+\sin ^{2}(\alpha )\sin
^{2}(\Theta )((\nabla \alpha )^{2}(D\Phi )^{2}-(\nabla \alpha \cdot D\Phi
)^{2}) + \notag \\
& +\sin ^{4}(\alpha )\sin ^{2}(\Theta )((\nabla \Theta )^{2}(D\Phi
)^{2}-(\nabla \Theta \cdot \nabla \Phi )^{2})\bigr)\biggl\}+I^{U(1)},
\label{actionexp1}
\end{align*}%
where 
\begin{equation*}
D_{\mu }\Phi =\nabla _{\mu }\Phi -2eA_{\mu }.
\end{equation*}%
It is a trivial computation to check that the covariant derivative in terms
of the three scalar degrees of freedom $\alpha (x^{\mu })$, $\Theta (x^{\mu
})$ and $\Phi (x^{\mu })$ is equivalent to the following minimal coupling
rule 
\begin{align*}
\nabla _{\mu }\alpha & \rightarrow D_{\mu }\alpha =\nabla_{\mu} \alpha , \\
\nabla _{\mu }\Theta & \rightarrow D_{\mu }\Theta =\nabla_{\mu} \Theta , \\
\nabla _{\mu }\Phi & \rightarrow D_{\mu }\Phi =\nabla_{\mu} \Phi -2eA_{\mu }.
\end{align*}%
Thus, the scalar degree of freedom $\Phi $ plays the role of the ``$U(1)$
phase" of the Skyrme field since, under a $U(1)$ gauge transformation, it
transforms as 
\begin{align*}
A_{\mu }& \rightarrow A_{\mu }+\nabla _{\mu }\Lambda , \\
\Phi & \rightarrow \Phi +2e\Lambda .
\end{align*}%
It is also useful to write the full Skyrme-Maxwell equations in terms of the
three scalar degrees of freedom $\alpha $, $\Theta $ and $\Phi $: 
\begin{align*}
& -\Box \alpha +\sin (\alpha )\cos (\alpha )\left( (\nabla \Theta )^{2}+\sin
^{2}(\Theta )(D\Phi )^{2}\right) +\lambda \Bigr(\sin (\alpha )\cos (\alpha )%
\bigr((\nabla \alpha )^{2}(\nabla \Theta )^{2}-(\nabla \alpha \cdot \nabla
\Theta )^{2}\bigr) +  \notag \\
& +\sin (\alpha )\cos (\alpha )\sin ^{2}(\Theta )\bigr((\nabla \alpha
)^{2}(D\Phi )^{2}-(\nabla \alpha \cdot D\Phi )^{2}\bigr)+2\sin ^{3}(\alpha
)\cos (\alpha )\sin ^{2}(\Theta )\bigr((\nabla \Theta )^{2}(D\Phi )^{2} -
\notag \\
& -(\nabla \Theta \cdot D\Phi )^{2}\bigr)-\nabla _{\mu }\bigr(\sin
^{2}(\alpha )(\nabla \Theta )^{2}\nabla ^{\mu }\alpha \bigr)+\nabla _{\mu }%
\bigr(\sin ^{2}(\alpha )(\nabla \alpha \cdot \nabla \Theta )\nabla ^{\mu
}\Theta \bigr) - \notag \\
& -\nabla _{\mu }\bigr(\sin ^{2}(\alpha )\sin ^{2}(\Theta )(D\Phi
)^{2}\nabla ^{\mu }\alpha \bigr)+\nabla _{\mu }\bigr(\sin ^{2}(\alpha )\sin
^{2}(\Theta )(\nabla \alpha \cdot D\Phi )D^{\mu }\Phi \bigr)\Bigr)=0
\end{align*}%
\begin{align*}
& -\sin ^{2}(\alpha )\Box \Theta -2\sin (\alpha )\cos (\alpha )(\nabla
\alpha \cdot \nabla \Theta )+\sin ^{2}(\alpha )\sin (\Theta )\cos (\Theta
)(D\Phi )^{2} + \notag \\
& +\lambda \Bigr(\sin ^{2}(\alpha )\sin (\Theta )\cos (\Theta )\bigr((\nabla
\alpha )^{2}(D\Phi )^{2}-(\nabla \alpha \cdot D\Phi )^{2}\bigr) + \notag \\
& +\sin ^{4}(\alpha )\sin (\Theta )\cos (\Theta )\bigr((\nabla \Theta
)^{2}(D\Phi )^{2}-(\nabla \Theta \cdot D\Phi )^{2}\bigr)-\nabla _{\mu }\bigr(%
\sin ^{2}(\alpha )(\nabla \alpha )^{2}\nabla ^{\mu }\Theta \bigr) -  \notag \\
& -\nabla _{\mu }\bigr(\sin ^{4}(\alpha )\sin ^{2}(\Theta )(D\Phi
)^{2}\nabla ^{\mu }\Theta \bigr)+\nabla _{\mu }\bigr(\sin ^{4}(\alpha )\sin
^{2}(\Theta )(\nabla \Theta \cdot D\Phi )D^{\mu }\Phi \bigr) +  \notag \\
& +\nabla _{\mu }\bigr(\sin ^{2}(\alpha )(\nabla \alpha \cdot \nabla \Theta
)\nabla ^{\mu }\alpha \bigr)\Bigr)=0\ ,  
\end{align*}%
\begin{align*}
& -\sin ^{2}(\alpha )\sin ^{2}(\Theta )\bigr(\Box \Phi -2e\nabla _{\mu
}A^{\mu }\bigr)-2\sin (\alpha )\cos (\alpha )\sin ^{2}(\Theta )(\nabla
\alpha \cdot D\Phi ) - \notag \\
& -2\sin ^{2}(\alpha )\sin (\Theta )\cos (\Theta )(\nabla \Theta \cdot D\Phi
)+\lambda \Bigr(-\nabla _{\mu }\bigr(\sin ^{2}(\alpha )\sin ^{2}(\Theta
)(\nabla \alpha )^{2}D^{\mu }\Phi \bigr) + \notag \\
& +\nabla _{\mu }\bigr(\sin ^{2}(\alpha )\sin ^{2}(\Theta )(\nabla \alpha
\cdot D\Phi )\nabla ^{\mu }\alpha \bigr)-\nabla _{\mu }\bigr(\sin
^{4}(\alpha )\sin ^{2}(\Theta )(\nabla \Theta )^{2}D^{\mu }\Phi \bigr) +
\notag \\
& +\nabla _{\mu }\bigr(\sin ^{4}(\alpha )\sin ^{2}(\Theta )(\nabla \Theta
\cdot D\Phi )\nabla ^{\mu }\Theta \bigr)\Bigr)=0\ .  
\end{align*}

\subsubsection{Euler angles parametrization}

For the $SU(2)$ case we chose the following element $U$ of $SU(2)$ 
\begin{equation}
U=\exp \left( F\left( x^{\mu }\right) t_{3}\right) \exp \left( H\left(
x^{\mu }\right) t_{2}\right) \exp \left( G\left( x^{\mu }\right)
t_{3}\right) ,  \label{I2}
\end{equation}%
where $F\left( x^{\mu }\right) $, $G\left( x^{\mu }\right) $ and $H\left(
x^{\mu }\right) $ are the three scalar degrees of freedom (traditionally, in
this parametrization the field $H$ is called profile). It is worth
emphasizing that, as it happens for the exponential representation discussed
in the previous subsection, any group element can always be written in the
Euler angle representation. We will see that, as has been already emphasized,
in order to construct a concrete ansatz that can be adapted to the finite
density situation it is convenient to write explicitly the Skyrme action in
terms of the Euler angle parametrization in Eq. (\ref{I2}). A direct
computation shows that 
\begin{align*}
I(H,F,G)=& -\frac{K}{2}\int d^{4}v\text{Tr}\biggl\{(\nabla
H)^{2}+(DF)^{2}+(DG)^{2}+2\cos (2H)(DF\cdot DG) - \notag \\
& -\lambda \bigr(2\cos (2H)((\nabla H\cdot DF)(\nabla H\cdot DG)-(\nabla
H)^{2}(DF\cdot DG)) + \notag \\
& +4\sin ^{2}(H)\cos ^{2}(H)((DF\cdot DG)^{2}-(DF)^{2}(DG)^{2}) + \notag \\
& +(\nabla H\cdot DF)^{2}+(\nabla H\cdot DG)^{2}-(\nabla
H)^{2}(DF)^{2}-(\nabla H)^{2}(DG)^{2}\bigr)\biggl\}\ .
\label{hadronicactionFFH}
\end{align*}%
The minimal coupling with the Maxwell field in this parametrization is 
\begin{equation*}
D_{\mu }F=\nabla _{\mu }F-eA_{\mu }\ ,\qquad D_{\mu }G=\nabla _{\mu
}G+eA_{\mu }\ .
\end{equation*}%
The field equations in the Euler parameterization are given by 
\begin{align}
0=& \square H+2\sin (2H)(\nabla F\cdot \nabla G)-\lambda \biggl\{2\sin (2H)%
\Bigr((\nabla H\cdot \nabla F)(\nabla H\cdot \nabla G)-(\nabla H)^{2}(\nabla
F\cdot \nabla G) -  \notag \\
& -\cos (2H)\bigr((\nabla F\cdot \nabla G)^{2}-(\nabla F)^{2}(\nabla G)^{2}%
\bigr)\Bigr)+\nabla _{\mu }(\cos (2H)(\nabla H\cdot \nabla G)\nabla ^{\mu }F) +
\notag \\
& +\nabla _{\mu }(\cos (2H)(\nabla H\cdot \nabla F)\nabla ^{\mu }G)-\nabla
_{\mu }(2\cos (2H)(\nabla G\cdot \nabla F)\nabla ^{\mu }H)+\nabla _{\mu
}((\nabla H\cdot \nabla F)\nabla ^{\mu }F) + \notag \\
& +\nabla _{\mu }((\nabla H\cdot \nabla G)\nabla ^{\mu }G)-\nabla _{\mu
}((\nabla F)^{2}\nabla ^{\mu }H)-\nabla _{\mu }((\nabla G)^{2}\nabla ^{\mu
}H)\biggl\}\ ,  
\label{fqomH}
\end{align}%
\begin{align}
0=& \square F-e\nabla \cdot A-2\sin (2H)(DG\cdot \nabla H)+\cos (2H)(\square
G+e\nabla \cdot A) -  \notag \\
& -\lambda \Bigr(\nabla _{\mu }(\cos (2H)(\nabla H\cdot DG)\nabla ^{\mu
}H)-\nabla _{\mu }(\cos (2H)(\nabla H)^{2}D^{\mu }G) +  \notag \\
& +\nabla _{\mu }(4\sin ^{2}(H)\cos ^{2}(H)(DF\cdot DG)D^{\mu }G)-\nabla
_{\mu }(4\sin ^{2}(H)\cos ^{2}(H)(DG)^{2}D^{\mu }F) +  \notag \\
& +\nabla _{\mu }((\nabla H\cdot DF)\nabla ^{\mu }H)-\nabla _{\mu }((\nabla
H)^{2}D^{\mu }F)\Bigr)\ ,  
\end{align}%
\begin{align}
0=& \square G+e\nabla \cdot A-2\sin (2H)(DF\cdot \nabla H)+\cos (2H)(\square
F-e\nabla \cdot A) - \notag \\
& -\lambda \Bigr(\nabla _{\mu }(\cos (2H)(\nabla H\cdot DF)\nabla ^{\mu
}H)-\nabla _{\mu }(\cos (2H)(\nabla H)^{2}D^{\mu }F) + \notag \\
& +\nabla _{\mu }(4\sin ^{2}(H)\cos ^{2}(H)(DF\cdot DG)D^{\mu }F)-\nabla
_{\mu }(4\sin ^{2}(H)\cos ^{2}(H)(DF)^{2}D^{\mu }G) + \notag \\
& +\nabla _{\mu }((\nabla H\cdot DG)\nabla ^{\mu }H)-\nabla _{\mu }((\nabla
H)^{2}D^{\mu }G)\Bigr)\ .  \label{fqomF3}
\end{align}%
In the following sections, we will discuss first these field equations in
the ungauged case to show how and why effective low energy chiral conformal
degrees of freedom appear. 


\section{Skyrmions at Finite Density}

The first references analyzed the effects of forcing Skyrmions to live at finite Baryon density were \cite{aprox1,aprox4,SkCr2,SkCr3,SkCr4,SkCr5}, and \cite{WITTEN1981513}, where the variational construction and
classifications of ordered arrays of Skyrmions were analyzed. In these
references, roughly speaking, each peak of the energy and Baryon densities
represent a single unit charge Skyrmion (Baryon). In the present review we
are more interested in situations where these peaks (representing charge-1
Skyrmions) can melt into bigger structures with much higher topological
charges (such as Hadronic layers and tubes).

\subsection{The first example: Sine-Gordon layer}

In \cite{56c}, the first topologically non-trivial analytical solution in flat space-time that describes Hadronic layers in a finite
volume was constructed. In order to confine the system to a flat region of
fine spatial volume in that reference the metric in 3 + 1 dimensions was
taken\footnote{%
In the following sections, we will see that actually the metric can be chosen
in a slightly more generic way.} to be
\begin{equation}
ds^{2}=-dt^{2}+L^{2}(dr^{2}+d\gamma ^{2}+d\phi ^{2}) \, ,
\end{equation}%
where the range of coordinates\footnote{%
The proper range of the spatial coordinate can be determined by taking into
account the theory of the Euler angle (see \cite{euler1,euler2,euler3}, and \cite{euler4}).} are 
\begin{equation*}
0\leq r\leq 2\pi \quad ,\quad 0\leq \gamma \leq 4\pi \quad ,\quad 0\leq \phi
\leq 2\pi .
\end{equation*}%
The ansatz for the Skyrme field was originally given in the exponential
parameterization in Eqs. (\ref{U1}) and (\ref{U2}) as follows
{\small
\begin{align}
\Phi & =\frac{\gamma +\phi }{2},  \notag \\
\tan {\Theta }& =\frac{\tan H}{\cos A},  \notag \\
\tan {\alpha }& =\frac{\sqrt{1+\tan ^{2}\Theta }}{\tan A} \, ,
\label{ch:Skf:eqscalarset}
\end{align}}%
where $A=(\gamma -\phi )/{2}$ and $H=H(t,r)$.

The remarkable property of the above ansatz is that it reduces the complete
set of field equations to a single equation, given by 
\begin{eqnarray}
\Box H-\frac{\lambda }{8L^{2}(\lambda +2L^{2})}\sin {(4H)} &=&0, \\
\frac{\partial ^{2}}{\partial t^{2}}-\frac{1}{L^{2}}\frac{\partial ^{2}}{%
\partial r^{2}} &=&\Box \, ,
\end{eqnarray}%
which is the Sine-Gordon equation, the prototype of integrable PDE!
Moreover, the topological charge is non-vanishing and it reads 
\begin{equation}
\rho _{B}=3\sin {(2H)}dHd\gamma d\phi .
\end{equation}%
Furthermore, when the profile $H$ satisfies the following boundary conditions%
\begin{equation}
H(t,0)=\ 0,\ \ H(t,2\pi )=\ \pm \frac{\pi }{2},
\end{equation}%
the topological charge takes the values of $B=\pm 1$. These configurations
represent the first analytic examples of a Baryonic layer in a flat box of
finite volume. In the following sections, we will discuss the generalizations
of these solutions.


\section{Modulated condensates and the appearance of chiral degrees of
freedom}

In this section, we present the explicit construction of modulated
inhomogeneous Hadronic condensates in the ungauged case. These modulations
are encoded in emergent chiral conformal degrees of freedom.

\subsection{Finite Density: Metric of a Box}

\label{sec:finitedensity} We are interested in analyzing the intriguing
phenomena that occur when a finite amount of Baryonic charge lives within a
finite spatial volume. We consider as a starting point a slightly more
general metric than the one introduced in the previous section (which allows
to describe a flat box with different sizes along the three spatial
directions $\{r,\theta ,\phi \}$): the line element is 
\begin{equation}
ds^{2}=-dt^{2}+L_{r}^{2}dr^{2}+L_{\theta }^{2}d\theta ^{2}+L_{\phi
}^{2}d\phi ^{2}\ ,  \label{Box}
\end{equation}%
where $L_{i}$ are constants representing the length of the box where the
solitons are confined. The adimensional coordinates $\{r,\theta ,\phi \}$
have the following ranges 
\begin{equation}
0\leq r\leq 2\pi \ ,\quad 0\leq \theta \leq \pi \ ,\quad 0\leq \phi \leq
2\pi \ ,  \label{ranges}
\end{equation}%
so that the volume available for the solitons is $V=4\pi ^{3}L_{r}L_{\theta
}L_{\phi }$. Notice that the coordinates $\{r,\theta ,\phi \}$ are
Cartesian, their finite ranges can be fixed for instance using the theory of
Euler angles \cite{euler1,euler2,euler3}, and \cite{euler4}. 

\subsection{Euler Angles parametrization: Hadronic layers}

\label{sec:our:skeuler}

The energy-momentum tensor of the Skyrme model \eqref{TmunuSK}, in the Euler angle
parametrization reads 
\begin{align*}
T_{\mu \nu }^{SK}& =g_{\mu \nu }\mathcal{L}^{SK}+K\Bigr\{\nabla _{\mu
}H\nabla _{\nu }H+D_{\mu }FD_{\nu }F+D_{\mu }GD_{\nu }G+\cos {(2H)}(D_{\mu
}FD_{\nu }G+D_{\nu }FD_{\mu }G) -  \notag \\
& -\lambda \Bigr(\cos {(2H)}\bigr((\nabla H\cdot DG)(\nabla _{\mu }HD_{\nu
}F+\nabla _{\nu }HD_{\mu }F)+(\nabla H\cdot DF)(\nabla _{\mu }HD_{\nu
}G+\nabla _{\nu }HD_{\mu }G) -  \notag \\
& -2(\nabla F\cdot DG)\nabla _{\mu }H\nabla _{\nu }H-(\nabla H)^{2}(D_{\mu
}FD_{\nu }G+D_{\nu }FD_{\mu }G)\bigr)+  \notag \\
& +4\sin ^{2}{H}\cos ^{2}H\bigr((DF\cdot DG)(D_{\mu }FD_{\nu }G+D_{\nu
}FD_{\mu }G)-(DG)^{2}\nabla _{\mu }F\nabla _{\nu }F-(DG)^{2}\nabla _{\mu
}G\nabla _{\nu }F\bigr)+  \notag \\
& +(\nabla H\cdot DF)(\nabla _{\mu }HD_{\nu }F+\nabla _{\nu }HD_{\mu
}F)+(\nabla H\cdot DG)(\nabla _{\mu }HD_{\nu }G+\nabla _{\nu }HD_{\mu }G)- 
\notag \\
& -\nabla _{\mu }H\nabla _{\nu }H((DF)^{2}+(DG)^{2})-(\nabla H)^{2}(D_{\mu
}FD_{\nu }F+D_{\mu }GD_{\nu }G)\Bigr)\Bigr\}\ .
\end{align*}%
The analysis of the energy-momentum tensor is relevant to determine the
contour plots and the shape of the condensates.

The ungauged configurations which represent Hadronic layers of arbitrary
Baryonic charge can be generalized by including an arbitrary light-like
function as a degree of freedom. The suitable generalization of the ansatz
in \cite{gaugsksu(n),56a0,56a,56a1,56a2,56b1}, and \cite{56c} is 
\begin{equation}
H=H(r)\ ,\quad F=q\theta \ ,\quad G=G(u)\ ,\qquad u=\frac{t}{L_{\phi }}-\phi
\ ,  \label{eq:G:lasagna}
\end{equation}%
where $q$ is an integer number. Here $G\left( u\right) $ is an arbitrary
function of the light-like coordinate $u$. The Eq. (\ref{eq:G:lasagna})
explicitly avoids the no-go Derrick's theorem \cite{Derrick} due to the time
dependence of the $U$ field. The above ansatz keeps all the nice properties
of the one in \cite{gaugsksu(n)}. In particular, it satisfies the following
identities 
\begin{equation}
(\nabla F\cdot \nabla G)=(\nabla H\cdot \nabla F)=(\nabla H\cdot \nabla
G)=(\nabla F)^{2}=0,  \label{usefulansatzdecouplasagna}
\end{equation}%
which greatly simplifies the field equations. In fact, the ansatz in Eq. %
\eqref{eq:G:lasagna} reduces the Skyrme field equations in Eqs. \eqref{fqomH}
and \eqref{fqomF3} to a simple linear equation 
\begin{equation*}
\partial _{r}^{2}H(r)=0\quad \Rightarrow H(r)=\kappa r+\kappa _{0}\ ,
\end{equation*}%
where $\kappa _{0}$ can be fixed to zero, the integration constant $\kappa $
will be determined using appropriate boundary conditions.

Plugging the ansatz in Eqs. \eqref{I2} , \eqref{eq:G:lasagna} into Eq. \eqref{Bcurrent}, the topological charge density of the matter field reads 
\begin{equation*}
\rho_{B}=-12q\sin (2H)H^{\prime }\partial _{\phi }G,
\end{equation*}%
where it can be seen that the appropriate boundary conditions for the
soliton profile $H(r)$ and the light-like function $G(u)$ are the following: 
\begin{equation}
H(r=2\pi )=\frac{\pi }{2}\ ,\quad H(r=0)=0\ ,\quad G(t,\phi =0)-G(t,\phi
=2\pi )=(2\pi )p\ ,  \label{BCL1}
\end{equation}%
so that the topological charge takes the value 
\begin{equation}
B=pq\ .  \label{BCL2}
\end{equation}%
Therefore, using the above boundary conditions, the profile becomes 
\begin{equation*}
H(r)=\frac{r}{4}.
\end{equation*}

\subsubsection{Physical interpretation of the chiral degrees of freedom}

In order to clarify the physical meaning of the function $G(u)$ appearing in
the ansatz in Eq. (\ref{eq:G:lasagna}), let us consider the slightly
different ansatz 
\begin{equation*}
H=\frac{r}{4}\ ,\quad F=q\theta \ ,\quad G=G(t,\phi )\ .
\end{equation*}%
With the ansatz here above the Skyrme field equations would reduce to%
\begin{equation*}
\left( \left( \frac{\partial }{\partial t}-\frac{1}{L_{\phi }}\frac{\partial 
}{\partial \phi }\right) G\right) \left( \left( \frac{\partial }{\partial t}+%
\frac{1}{L_{\phi }}\frac{\partial }{\partial \phi }\right) G\right) =0.
\end{equation*}%
Thus, the Skyrme field equations force the choice of a chirality: $G$ can
represent either left movers or a right movers (but cannot represent both).
Let us choose, as in Eq. (\ref{eq:G:lasagna}) $G=G(u)$. Then, the boundary
conditions in Eqs. (\ref{BCL1}) and (\ref{BCL2}) ensure that $G(u)$ has the
following expression:%
\begin{equation}
G\left( u\right) =pu+\widetilde{G}\left( u\right) \ ,  \label{example2}
\end{equation}%
where $\widetilde{G}\left( u\right) $ is periodic in the coordinate $\phi $:%
\begin{equation*}
\widetilde{G}\left( u\right) =\sum_{N}a_{N}\cos \left( Nu\right) +b_{N}\sin
\left( Nu\right) \ ,\ N\in 
\mathbb{N}
\ ,
\end{equation*}%
where $a_{N}$ and $b_{N}$ are real coefficients and the integer $N$ labels
the chiral modes. Thus, the first term (linear in $u$) on the right hand
side of Eq. (\ref{example2}) contributes to the Baryonic charge while $%
\widetilde{G}$ does not (being periodic in the coordinate $\phi $). In order
to interpret $\widetilde{G}$ it is enough to observe that, when $\widetilde{G%
}=0$, the energy-momentum tensor only depends on the coordinate $r$ (while
it is homogeneous in the other two spatial coordinates). 
\begin{figure}[h]
\centering
\includegraphics[width=1.\textwidth]{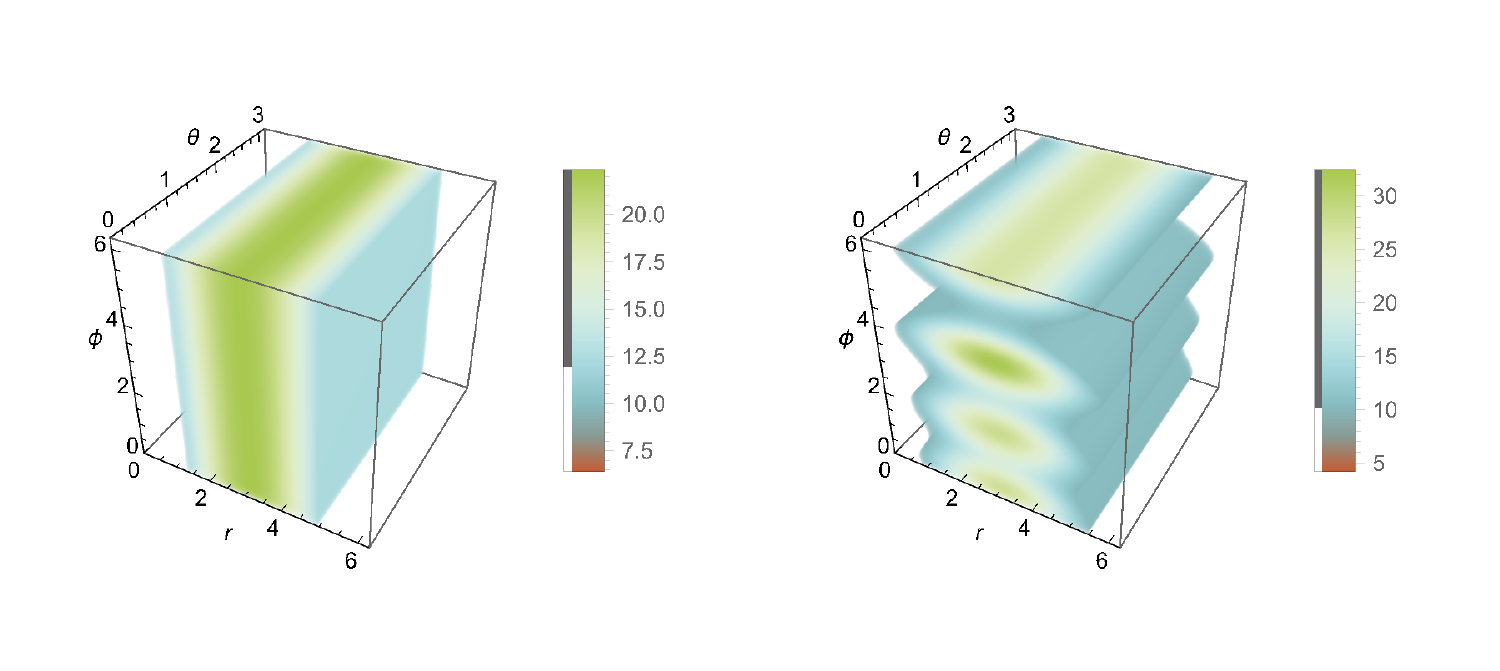}  
\caption{Energy density with and without modulation of hadronic layers, with
topological charge $B=4$. For both cases we have set $L_{r}=L_{\protect%
\theta }=L_{\protect\phi }=K=\protect\lambda =1$ and $p=q=2$. When $%
a_{I}=b_{I}=0$, we obtain the left plot for hadronic layers without
modulations. However, if we consider $a_{1}=a_{2}=b_{1}=-b_{3}=0.1$, we
obtain the right plot for hadronic layers with modulations.}
\label{fig:lasagna}
\end{figure}
Thus, in this case the solution represents a homogeneous Baryonic layer. On
the other hand, when we turn on $\widetilde{G}$, the energy-momentum tensor
depends not only on $r$ but also on $u$ (through the modes of $\widetilde{G}$%
). In this case, the plots \ref{fig:lasagna} of the energy-density reveal
that $\widetilde{G}$ represents modulations of the layer in the $\phi $
direction moving at the speed of light. Consequently, the present family of
exact analytic solutions of the Skyrme field equations represents Baryonic
layers dressed by modulations in the $\phi $ direction. 

It is worth emphasizing here the high theoretical interest to reveal the
emergence of chiral conformal degrees of freedom living on Hadronic layers.
Such chiral conformal degrees of freedom are not only exact solutions of the
full Skyrme field equations (as described in detail here above): such chiral
field $\widetilde{G}$ can be also considered as a solution of the linearized
Skyrme field equations on top of the exact solution where $\widetilde{G}=0$
(which represents a homogeneous Baryonic layer). This observation allows to
interpret $\widetilde{G}$ as a chiral field propagating within a Baryonic
medium (provided by the homogeneous Baryonic layer itself). Hence, it is
interesting to analyze the transport properties of such chiral field (as
such analysis could reveal how the Baryonic medium affects this important
quantities): we hope to come back on this issue in a future publication.

\subsection{Exponential parametrization: Hadronic tubes}

\label{sec:our:skhed} The energy-momentum tensor of the Skyrme Model \eqref{TmunuSK}, in the exponential parametrization \eqref{U1} with \eqref{U2} reads 
\begin{align*}
T_{\mu \nu }& =g_{\mu \nu }\mathcal{L}^{\text{SK}}-K\Bigr\{\nabla _{\mu
}\alpha \nabla _{\nu }\alpha +\sin ^{2}\!\alpha \left( \nabla _{\mu }\Theta
\nabla _{\nu }\Theta +\sin ^{2}\!\Theta \,D_{\mu }\Phi D_{\nu }\Phi \right)
\ + \\
& +\ \lambda \Bigr[\sin ^{2}\!\alpha \left( \left( \nabla \Theta \right)
^{2}\nabla _{\mu }\alpha \nabla _{\nu }\alpha +\left( \nabla \alpha \right)
^{2}\nabla _{\mu }\Theta \nabla _{\nu }\Theta -\left( \nabla \alpha \!\cdot
\!\nabla \Theta \right) (\nabla _{\mu }\alpha \nabla _{\nu }\Theta +\nabla
_{\nu }\alpha \nabla _{\mu }\Theta )\right) \ +  \notag \\
& +\sin ^{2}\!\alpha \,\sin ^{2}\Theta \left( \left( D\Phi \right)
^{2}\nabla _{\mu }\alpha \nabla _{\nu }\alpha +\left( \nabla \alpha \right)
^{2}D_{\mu }\Phi D_{\nu }\Phi -\left( \nabla \alpha \!\cdot \!D\Phi \right)
(\nabla _{\mu }\alpha D_{\nu }\Phi +\nabla _{\nu }\alpha D_{\mu }\Phi
)\right) \ +  \notag \\
& +\sin ^{4}\!\alpha \sin ^{2}\!\Theta \left( \left( D\Phi \right)
^{2}\nabla _{\mu }\Theta \nabla _{\nu }\Theta +\left( \nabla \Theta \right)
^{2}D_{\mu }\Phi D_{\nu }\Phi -\left( \nabla \Theta \cdot D\Phi \right)
(\nabla _{\mu }\Theta D_{\nu }\Phi +\nabla _{\nu }\Theta D_{\mu }\Phi
)\right) \Bigr]\Bigr\}\ ,  \notag
\end{align*}
$\mathcal{L}^{\text{SK}}$ being the Skyrme Lagrangian, while the topological
charge density in Eq. \eqref{Bcurrent} becomes 
\begin{equation*}
\rho_{B}=12\sin ^{2}\!\alpha \sin \Theta d\Phi \wedge d\Theta \wedge
d\alpha \ .  \label{B01}
\end{equation*}%
From the above expression, it follows that to have non-trivial topological
configurations, we must demand that $d\Theta \wedge d\Phi \wedge d\alpha
\neq 0$. This implies the necessary (but insufficient) condition that $\alpha $, $\Theta $, and $\Phi $
must be three independent functions in order
for the Baryon charge to be non-zero.

The suitable generalization of the ansatz in \cite{crystal1} and \cite%
{crystal2} which (as in the case of Hadronic layers) allows to generalize
the analytic configurations representing Hadronic tubes to inhomogeneous
tubes (namely, Hadronic tubes which are not anymore homogeneous along their
axis) is 
\begin{align}
\alpha & =\alpha (r),  \notag \\
\Theta & =Q\theta ,\hspace{0.2cm}\quad \quad Q=2v+1\ ,\ \ v\in \mathbb{Z}, 
\notag \\
\Phi & =G\left( u\right) ,\quad u=\frac{t}{L_{\phi }}-\phi ,
\label{Uexplicit}
\end{align}%
where now $G\left( u\right) $ is an arbitrary function of the light-like
coordinate $u$. It is easy to see that the above ansatz keeps all the nice
properties of the one in \cite{crystal1} and \cite{crystal2}. Firstly, it
satisfies the following identities 
\begin{equation}
(\nabla \Phi \cdot \nabla \alpha )=(\nabla \alpha \cdot \nabla \Theta
)=(\nabla \Theta \cdot \nabla \Phi )=(\nabla \Phi )^{2}=0\ .
\label{fulansatzdecoupling}
\end{equation}%
The great usefulness of the above identities is that they allow to decouple the Skyrme field equations for the three degrees of freedom ($\alpha $, $%
\Phi $ and $\Theta $) without killing the topological charge. With the
ansatz introduced in Eqs. (\ref{U1}), (\ref{U2}) and (\ref{Uexplicit}) the
Skyrme field equations reduce to a second order ODE for the profile $\alpha $%
: 
\begin{equation}
\alpha ^{\prime \prime }+\frac{Q^{2}\sin (\alpha )\cos (\alpha )(\lambda
\alpha ^{\prime 2}-L_{r}^{2})}{L_{\theta }^{2}+\lambda Q^{2}\sin ^{2}(\alpha
)}=0\ ,  \label{etaE0}
\end{equation}%
which can be reduced to a first-order ODE that can be conveniently written
as 
\begin{equation}
\frac{d\alpha }{\eta (\alpha ,E_{0})}=\pm dr\ ,\quad \eta \left( \alpha
,E_{0}\right) =\biggl[\frac{E_{0}L_{\theta }^{2}-\frac{1}{2}%
q^{2}L_{r}^{2}\cos (2\alpha )}{L_{\theta }^{2}+\lambda Q^{2}\sin ^{2}(\alpha
)}\biggl]^{\frac{1}{2}}\ ,  \label{alpha}
\end{equation}%
$E_{0}$ being an integration constant (fixed by the boundary conditions, as
we will see below)\footnote{%
Eq. \eqref{alpha} can be solved analytically in terms of Elliptic Functions;
however, the explicit solution is not necessary for our purposes.}.

Plugging the ansatz in Eqs. (\ref{U1}), (\ref{U2}) and (\ref{Uexplicit})
into Eq. \eqref{Bcurrent}, the topological charge turns out to be 
\begin{equation*}
\rho_{B}= 12 q \sin(q\theta) \sin^2(\alpha) \alpha^{\prime }\partial _{\phi
}G\ ,
\end{equation*}
where it is clearly seen that the appropriate boundary conditions for the
soliton profile $\alpha (r)$ and the light-like function $G(u)$ are the
following: 
\begin{align}
\alpha (2\pi )-\alpha (0) &= n\pi, \notag \\
G(t,\phi=2\pi)-G(t,\phi=0) &= (2 \pi) p,  \label{bc}
\end{align}
with $n$ and $p$ integer numbers. Therefore, using Eq. \eqref{bc} and
integrating with the ranges defined in Eq. \eqref{ranges}, the topological
charge takes the value 
\begin{equation*}
B= n p \, .
\end{equation*}
We have used that $q$ is an odd number, as specified in the ansatz in Eq. %
\eqref{Uexplicit}. From Eqs. \eqref{ranges}, \eqref{alpha} and \eqref{bc} it
follows that the integration constant $E_{0}$ must satisfy 
\begin{equation}
n\int_{0}^{\pi }\frac{d\alpha }{\eta \left( \alpha ,E_{0}\right) }=2\pi \ .
\label{E0n}
\end{equation}
Eq. \eqref{E0n} is an equation for $E_{0}$ in terms of $n$ that always has a
real solution when 
\begin{equation*}
E_{0}>\frac{Q^2 L_r^2}{ 2L_\theta^2} \ ,
\end{equation*}
so that, for given values of $q$, $L_r$ and $L_\theta$, the integration
constant $E_{0}$ determines the value of the $\alpha $ profile for the
boundary conditions defined in Eq. \eqref{bc}.

From the above condition, it is clear that for large $n$, the integration
constant $E_{0}$ scales as $n^2 $ 
\begin{align*}
E_{0} &= n^2 \xi_{0}, \quad \quad \xi_{0} > 0,  
\end{align*}
where $\xi_{0}$ can also be interpreted as an integration constant and does
not depend on $n$ for large $n$.

\subsubsection{Physical interpretation of the chiral degrees of freedom}

Additionally, in the present case one can clarify the physical meaning of
the function $G(u)$ appearing in the ansatz in Eq. (\ref{Uexplicit}) by the
slightly more general ansatz 
\begin{align*}
\alpha & =\alpha (r),\ \frac{d\alpha }{\eta (\alpha ,E_{0})}=\pm dr\ , 
\notag \\
\quad \eta \left( \alpha ,E_{0}\right) & =\biggl[\frac{E_{0}L_{\theta }^{2}-%
\frac{1}{2}q^{2}L_{r}^{2}\cos (2\alpha )}{L_{\theta }^{2}+\lambda Q^{2}\sin
^{2}(\alpha )}\biggl]^{\frac{1}{2}}\ ,  \notag \\
\Theta & =Q\theta ,\hspace{0.2cm}\quad \quad Q=2v+1\ ,\ \ v\in \mathbb{Z}, 
\notag \\
\Phi & =G\left( t,\phi \right) ,\quad \alpha (2\pi )-\alpha (0)=n\pi \ , 
\end{align*}%
where $\alpha (r)$ and $\Theta (\theta )$ are the same as in Eq. (\ref%
{Uexplicit}) but $G$ has been taken as a generic function of $t$ and $\phi $
(instead of taking $G$ as a function of a single light-like coordinate). In
this way one can shed light on the true nature of $G$. As in the case of the
Baryonic layers described in the previous sections, with the ansatz here
above the Skyrme field equations reduce to%
\begin{equation}
\left( \left( \frac{\partial }{\partial t}-\frac{1}{L_{\phi }}\frac{\partial 
}{\partial \phi }\right) G\right) \left( \left( \frac{\partial }{\partial t}+%
\frac{1}{L_{\phi }}\frac{\partial }{\partial \phi }\right) G\right) =0\ ,
\label{example1S1.1}
\end{equation}%
plus%
\begin{equation*}
\left( \frac{\partial ^{2}}{\partial t^{2}}-\frac{1}{L_{\phi }^{2}}\frac{%
\partial ^{2}}{\partial \phi ^{2}}\right) G=0 \, ,
\end{equation*}%
which is a consequence of Eq. (\ref{example1S1.1}). Thus, $G$ can represent
either left movers or right movers (but cannot represent both). Let us then
choose $G=G(u)$. The boundary conditions in Eq. (\ref{bc}) require that $G(u)
$ has the following expression:%
\begin{equation}
G\left( u\right) =pu+\widetilde{G}\left( u\right) \ ,  \label{example2S}
\end{equation}%
where $\widetilde{G}\left( u\right) $ is periodic in the coordinate $\phi $:%
\begin{equation*}
\widetilde{G}\left( u\right) =\sum_{N}a_{N}\cos \left( Nu\right) +b_{N}\sin
\left( Nu\right) \ ,\ N\in 
\mathbb{N}
\ ,
\end{equation*}
\begin{figure}[h!]
\centering
\includegraphics[width=1.\textwidth]{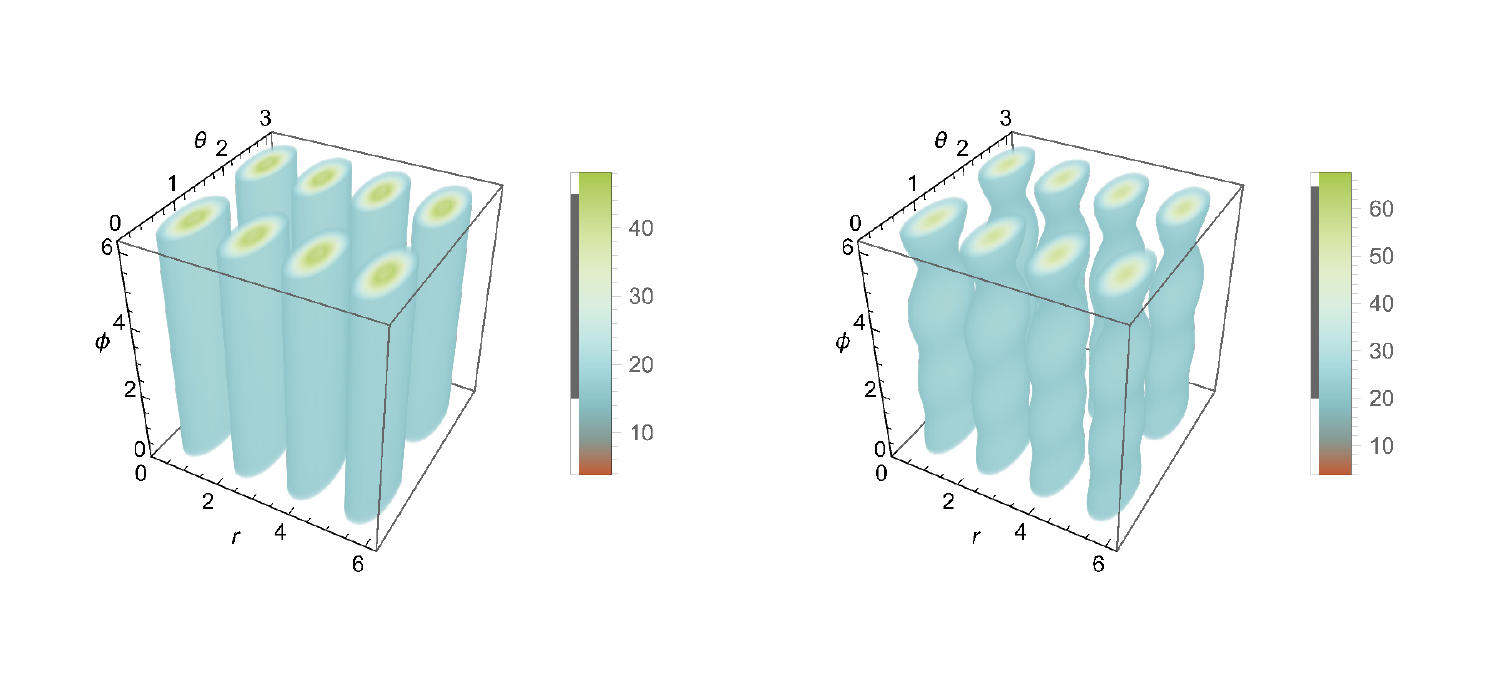}  
\caption{Energy density with and without modulation of hadronic tubes, with
topological charge $B=8$. For both cases we have set $L_r = L_{\protect\theta%
} = L_{\protect\phi} = K = \protect\lambda = 1$, $p = Q = 2$ and $n=4$. When 
$a_{I}= b_{I} = 0$ we obtain the left plot for hadronic tubes without
modulations. However, if we consider $a_{1} = a_{2} = b_{1} = -b_{3} =0.1$
we obtain the right plot for hadronic tubes with modulations.}
\label{fig:spaghetti}
\end{figure}
where $a_{N}$ and $b_{N}$ are real coefficients. Furthermore, in this case
the first term (linear in $u$) on the right hand side of Eq. (\ref{example2S}%
) contributes to the Baryonic charge while $\widetilde{G}$ does not (being
periodic in the coordinate $\phi $). Moreover, when $\widetilde{G}=0$, the
stationary energy-momentum tensor only depends on the coordinates $r$ and $%
\theta $ (while it does not depend on $\phi $).

Thus, in this case the solution represents an ordered array of homogeneous
Baryonic tubes. On the other hand, when we turn on $\widetilde{G}$, the
energy-momentum tensor depends not only on $r$ and $\theta $ but also on $u$
(through the modes of $\widetilde{G}$). In this case, the plots \ref%
{fig:spaghetti} of the energy-density reveal that $\widetilde{G}$ describes
modulations of the tubes in the $\phi $ direction moving at the speed of
light. These analytic solutions are Baryonic tubes dressed by modulations in
the $\phi $ direction.

Also in this case, the emergence of chiral conformal degrees of freedom
living on Hadronic tubes is a quite remarkable phenomenon. As in the case of
the Hadronic layers, such chiral conformal degrees of freedom are not only
exact solutions of the full Skyrme field equations: $\widetilde{G}$ can be
also considered as a solution of the linearized Skyrme field equations on
top of the exact solution where $\widetilde{G}=0$ (which represents ordered
arrays of homogeneous Baryonic tubes). This observation allows to interpret $%
\widetilde{G}$ as a chiral field propagating within the corresponding
Baryonic medium. Hence, it is interesting to analyze the transport properties of such a chiral field (as such analysis could reveal how the Baryonic medium affects these important quantities): we hope to come back on
this issue in a future publication.

\section{Gauged Skyrmions and applications}
In this section, we will discuss how to generalize the modulated inhomogeneous Baryonic condensates constructed in the previous sections in the case in which the minimal coupling with Maxwell theory is taken into account. This issue is extremely important for the following reason. The numerical simulations discussing the appearance of inhomogeneous Baryonic condensates (see \cite%
{pasta1,pasta2,pasta2a,pasta2b,pasta3,pasta4,pasta5,pastacond4,pasta7,pasta8,pasta9}%
, and the nice up to date review \cite{pasta10}) usually do not consider explicitly the electromagnetic interactions of the Baryons (which often are modeled as point-like particles). The self-consistent coupling of many Baryons with the electromagnetic field generated by the Baryons themselves
would make the numerical simulations considerably heavier. Therefore, to
have analytic tools which can help to understand the nature of the
electromagnetic field naturally associated to these inhomogeneous Hadronic
condensates would be extremely helpful also as a guide for numerical
simulations. Here we will describe how one can 
\"{}%
the condensates introduced in the previous sections with their own
electromagnetic fields. We will try to give a unified description of this
construction which is valid both for Hadronic layers and tubes.

The starting point is the observation that the key property of the ansatz in the ungauged cases which allows decoupling the Skyrme field equations without killing the topological charge is the relations in Eqs. (\ref%
{usefulansatzdecouplasagna}) and (\ref{fulansatzdecoupling}) for layers and tubes respectively. Thus, we have to construct an ansatz for the gauge potential $A_{\mu }$ which keeps (the covariant version of) Eqs. (\ref%
{usefulansatzdecouplasagna}) and (\ref{fulansatzdecoupling})\ alive. The answer to this question was found in \cite{crystal2,crystal3,56a}, and \cite%
{56a1}.

We will first describe the method in the case of the Hadronic tube, which is easier to understand. In order to minimally couple the Hadronic tubes discussed in the previous sections with $U(1)$ Maxwell gauge field let us consider a gauge potential $A_{\mu }$ with the following characteristics: 
\begin{equation}
A_{\mu }\partial ^{\mu }\alpha =0,\,A_{\mu }\partial ^{\mu }\Theta
=0,\,A_{\mu }\partial ^{\mu }G=0\ ,  \label{auxMT1}
\end{equation}%
\begin{equation}
A_{\mu }A^{\mu }=0\ ,\ \ \partial _{\mu }A^{\mu }=0\ .  \label{auxMT2}
\end{equation}
\subsection{Abelian Higgs Model}

Here we will clarify the conditions in Eqs. (\ref{auxMT1}) and (\ref{auxMT2}%
) in a simpler case: the Abelian-Higgs model whose action $S_{AH}$ is:%
\begin{eqnarray*}
S_{AH} &=&-\frac{1}{2}\int d^{4}x\sqrt{-g}\left[ \left( D_{\mu }\Psi ^{\ast
}\right) D^{\mu }\Psi +\frac{\gamma }{2}\left( \left\vert \Psi \right\vert
^{2}-v^{2}\right) ^{2}+\frac{1}{2}F_{\mu \nu }F^{\mu \nu }\right] \ , \\
D_{\mu }\Psi  &=&\partial _{\mu }\Psi -ieA_{\mu }\Psi \ ,\ \ F_{\mu \nu
}=\partial _{\mu }A_{\nu }-\partial _{\nu }A_{\mu }\ ,\ \Psi =h\exp \left(
iG\right) \ ,
\end{eqnarray*}%
where $\Psi $ is a complex scalar field (which can be parametrized in terms
of two real scalar degrees of freedom: the amplitude $h(x^{\mu })$ and the
phase $G(x^{\mu })$). The field equations of the Abelian-Higgs model are:%
\begin{eqnarray}
-D_{\mu }D^{\mu }\Psi +\gamma \left( \left\vert \Psi \right\vert
^{2}-v^{2}\right) \Psi  &=&0  \label{AB1} \\
-\partial^{\mu }F_{\mu \nu }+\text{Im}\left( \Psi ^{\ast }D_{\nu }\Psi
-\Psi D_{\nu }\Psi ^{\ast }\right)  &=&0\ .  \label{AB2}
\end{eqnarray}%
In general, the above system of equations is quite complicated (so much so
that not even in the BPS limit one can find analytic vortex-like solutions:
see \cite{manton}\ and references therein). Nevertheless, one can change a
little bit point of view asking the following question: can we choose an
ansatz for the Higgs field $\Psi $ and for the Abelian field $A_{\mu }$ in
such a way to decouple the equation of the Higgs field in Eq. (\ref{AB1})
from the Maxwell gauge potential $A_{\mu }$ (keeping alive the field
strength $F_{\mu \nu }$)? In order to answer this question, let us make a
list of all the terms which couple the Higgs field $\Psi $ with $A_{\mu }$
in Eq. (\ref{AB1}): such terms are%
\begin{equation*}
A_{\mu }A^{\mu }\Psi \ ,\ \ A_{\mu }\partial ^{\mu }\Psi \ ,\ \ \Psi
\partial ^{\mu }A_{\mu }\ .
\end{equation*}%
Thus, if there are non-trivial configurations such that the above terms all
vanish then the gauge potential would disappear from the Higgs equation and
all the terms that couple the gauge potential to the Higgs field would be
left in the Maxwell equation Eq. (\ref{AB2}). This is a very useful
technical achievement since in this way one can solve first the Higgs field equation without worrying about the Maxwell field and then the Maxwell
equations can be solved using the Higgs field as input. Examples of
configurations with such properties in the Abelian-Higgs model have been
found in \cite{LastNiKo}. The key idea is to choose a gauge potential with
two components defining a light-like vector, so that $A_{\mu }A^{\mu }=0$
(keeping alive $F_{\mu \nu }$). Moreover, one can choose the phase $G$ to
depend on a light-like variable in such a way that $A_{\mu }\partial ^{\mu
}G=0$ and the amplitude $h(x^{\mu })$ can be chosen to depend on space-like
coordinates corresponding to the spatial directions which are absent in $%
A_{\mu }$ so that $A_{\mu }\partial ^{\mu }h=0$. Last but not least, the
spatial dependence of the gauge potential can be chosen so that $\partial
^{\mu }A_{\mu }=0$. An explicit example of where this decoupling strategy works
is%
\begin{eqnarray*}
ds^{2} &=&-dt^{2}+L_{r}^{2}dr^{2}+L_{\theta }^{2}d\theta ^{2}+L_{\phi
}^{2}d\phi ^{2}\ , \\
A_{\mu } &=&(\varpi ,0,0,-L_{\phi }\varpi )\ \qquad \varpi =\varpi (r,\theta
)\ ,\ u=\frac{t}{L_{\phi }}-\phi \ , \\
\Psi  &=&h\exp \left( iG\right) \ ,\ \ h=h(r,\theta )\ ,\ G=Pu\ ,
\end{eqnarray*}%
where $P$ is an integer. One can check directly that with the ansatz here
above Eq. (\ref{AB1}) reduces to just one PDE for $h$ (where the gauge
potential does not appear) while the Maxwell equations of the Abelian-Higgs
model reduce to just one linear equation for $\varpi $ where $h$ plays the
role of an effective Schrodinger-like potential. In conclusion, in the
Abelian Higgs model it is possible to decouple the Higgs equation from the Maxwell gauge potential by adopting the strategy explained here above. It is worth emphasize here that the requirements $A_{\mu }A^{\mu }=0$ and $A_{\mu
}\partial ^{\mu }\Psi =0$ \textit{are not gauge-fixing choices} (since we
are already requiring $\partial ^{\mu }A_{\mu }=0$ and it is not possible to
implement more than one gauge fixing condition at the same time). Instead,
the conditions $A_{\mu }A^{\mu }=0$ and $A_{\mu }\partial ^{\mu }\Psi =0$
must be seen as a guide to find the best possible ansatz which is able to
reduce the complicated field equations of the Abelian Higgs model to a
solvable system of equations in a consistent way (keeping alive the
interactions between the matter field and the gauge field). \textit{A priori}%
, one could think that to require the three conditions ($A_{\mu }A^{\mu }=0$%
, $A_{\mu }\partial ^{\mu }\Psi =0$ and  $\partial ^{\mu }A_{\mu }=0$)\ at the same time is too restrictive and only trivial configurations can satisfy
all of them. In fact, as it has been shown in \cite{LastNiKo}, this is not
the case and many examples can be explicitly constructed. The same strategy
also works in the more complicated case of the gauged Skyrme Maxwell system
as we will now discuss.

Let us now go back to the gauged Skyrme model and to the conditions in Eqs. (%
\ref{auxMT1}) and (\ref{auxMT2}). One observes that the above conditions for 
$A_{\mu }$ are not empty. In order to satisfy both conditions it is enough
to consider a gauge potential with components only along the $t-$direction
and the $\phi -$direction, then these two components, $A_{t}$ and $A_{\phi }$%
, must be proportional (in order to satisfy $A_{\mu }A^{\mu }=0$) and can
depend on $r$, $\theta $ and the same null coordinate $u$ which enters in $G$%
: 
\begin{gather}
A_{\mu }=(\xi ,0,0,-L_{\phi }\xi )\ \qquad \xi =\xi (u,r,\theta )\ ,
\label{omega} \\
\Rightarrow A_{\mu }A^{\mu }=0\quad \text{ and}\quad \nabla ^{\mu }A_{\mu
}=0\ .  \notag
\end{gather}%
One can easily check that the field strength of such gauge potential in
general is non-vanishing. The above conditions in Eqs. (\ref{auxMT1}) and (%
\ref{auxMT2}) complement the condition in Eq. (\ref{fulansatzdecoupling})
for the $SU(2)$-valued Skyrme field. From the viewpoint of the gauged Skyrme
field equations, Eqs. (\ref{auxMT1}) and (\ref{auxMT2}) possess the very
welcome feature to eliminate all the terms of the gauged Skyrme field
equations which could, potentially, mix the $SU(2)$ degrees of freedom with
the gauge potential. Consequently, with the above ansatz for the gauge
potential, the gauged Skyrme field equations remain the same as the ungauged
Skyrme field equations corresponding to the ansatz in Eq. (\ref{Uexplicit}).
On the other hand, one may wonder whether the above conditions in Eqs. (\ref%
{auxMT1}) and (\ref{auxMT2}) are too restrictive from the viewpoint of the
Maxwell equations. In particular, the left hand side of the Maxwell field
equations (namely $\partial ^{\mu }F_{\mu \nu }$) only have components along
the $t-$direction and the $\phi -$direction (due to the form of the gauge
potential in Eq. (\ref{omega})). Thus, we have to analyze the $U(1)$ Skyrme
current. A direct computation shows, that in the exponential
parametrization, the right hand side $J_{\mu }$ of the Maxwell equations %
\eqref{Eqomaxwell} is {\small 
\begin{equation*}
J_{\mu }=-eK\sin ^{2}\alpha \sin ^{2}\Theta \biggl\{D_{\mu }\Phi +\frac{%
\lambda }{2}\Bigr((\nabla \alpha )^{2}D_{\mu }\Phi -(\nabla \alpha \cdot
D\Phi )\nabla _{\mu }\alpha +\sin ^{2}\alpha \bigr((\nabla \Theta
)^{2}D_{\mu }\Phi -(\nabla \Theta \cdot D\Phi )\nabla _{\mu }\Theta \bigr)%
\Bigr)\biggl\}\ .
\end{equation*}%
The terms which could spoil the consistency of the ansatz are all the terms
which are not proportional to }$D_{\mu }\Phi $ (such as the terms
proportional to $\nabla _{\mu }\alpha $ and $\nabla _{\mu }\Theta $). In
fact, all these 
\"{}%
vanish (since the ansatz has the property that $\nabla \alpha \cdot D\Phi
=0=\nabla \Theta \cdot D\Phi $). Hence, quite remarkably, with the above
choice of the gauge potential the field equations of the gauged Skyrme
Maxwell theory reduce exactly (no approximation involved here) in a
consistent way to Eq. (\ref{etaE0}) for $\alpha $ and only one linear
Schrodinger-like equation $\xi $ (see \cite{crystal2} and \cite{crystal3}).
From the mathematical viewpoint, this ansatz for the gauge potential has
been chosen to simplify as much as possible the coupled gauged Skyrme
Maxwell system keeping alive the field strength and the interactions. From
the physical viewpoint it turns out that such gauge fields belong to the
important class of force free Maxwell field \cite{crystal4.1} (which are
very important in astrophysics: see \cite%
{forcefree1,forcefree2,forcefree3,forcefree4,fprcefree5,forcefree6}\ and
references therein). Hence, from the physical viewpoint, the present
approach disclosed a relevant property of the inhomogeneous condensates
introduced in the previous sections which would have been very difficult to
discover with other methods: such condensates are natural sources of force
free plasmas.

As far as the Hadronic layers are concerned, the story is very similar
although slightly more complicated (see \cite{56a,56a1}, and \cite{56a11}).
The ansatz for the Hadronic layers 
\begin{equation*}
H=H(r)\ ,\quad F=q\theta \ ,\quad G=G(u)\ ,\qquad u=\frac{t}{L_{\phi }}-\phi
\ ,
\end{equation*}%
which satisfies the useful identities 
\begin{equation*}
(\nabla F\cdot \nabla G)=(\nabla H\cdot \nabla F)=(\nabla H\cdot \nabla
G)=(\nabla F)^{2}=0 \, ,
\end{equation*}%
can also be complemented with a suitable gauge potential by requiring that
the gauge potential does not spoil the solvability of the gauged Skyrme
field equations. The only difference is that instead of requiring $A_{\mu
}A^{\mu }=0$ one has to require a suitable quadratic constraint on the
components of $A_{\mu }$ (see \cite{56a,56a1}, and \cite{56a11}) together
with the usual conditions on the orthogonality of the gradients as in the
case of Hadronic layers. Furthermore, final results are similar: it is
possible to explicitly construct an ansatz for the gauge potential which
allows the gauged Skyrme field equations to be solved and the Maxwell
equations with the $U(1)$ Skyrme current reduce consistently to a
Schrodinger-like equation. The case of Hadronic layers is simpler since the
Schrodinger-like equation can be reduced to the Mathieu equation which is
solvable in terms of special functions. For the Hadronic layers it is also
true that they are natural sources of force free plasmas (see \cite{56a11}\
for details).

\section{Applications to Yang-Mills-Higgs theory}

In this section, we will discuss how the previous results on the Skyrme
model can be used in the Yang-Mills-Higgs case to disclose the
existence of chiral conformal modes dressing topologically non-trivial
configurations (see \cite{modulation} and \cite{nonlinearfab}).

\subsection{Pure Yang-Mills theory}


The Yang-Mills theory in $(3+1)$-dimensions is described by the action 
\begin{eqnarray*}
I[A] &=&\frac{1}{2e^{2}}\int d^{4}x\sqrt{-g}\,\text{Tr}(F_{\mu \nu }F^{\mu
\nu })\ ,  \label{I} \\
F_{\mu \nu } &=&\partial _{\mu }A_{\nu }-\partial _{\nu }A_{\mu }+[A_{\mu
},A_{\nu }]\,,
\end{eqnarray*}%
where $e$ is the coupling constant and $F_{\mu \nu }$ is the field strength.
The field equations and the energy-momentum tensor read 
\begin{equation*}
\nabla _{\nu }F^{\mu \nu }+[A_{\nu },F^{\mu \nu }]=\ 0\ ,  \label{yme0}
\end{equation*}%
\begin{equation*}
T_{\mu \nu }=-\frac{2}{e^{2}}\text{Tr}\biggl(F_{\mu \alpha }F_{\nu
}{}^{\alpha }-\frac{1}{4}g_{\mu \nu }F_{\alpha \beta }F^{\alpha \beta }%
\biggl)\ .  \label{TmunuYM}
\end{equation*}

The experience with the Skyrme model suggests that a very effective way to
confine topologically non-trivial configurations to a flat box with finite
spatial volume is to use the flat metric defined in Eq.~\eqref{Box}, with
the ranges 
\begin{equation}
0\leq \theta \leq 2\pi \ ,\qquad 0\leq \phi \leq \pi \ ,\qquad 0\leq r\leq
4\pi \ .\   \label{ranges1}
\end{equation}%
The above ranges for the coordinates $\theta $, $\phi $ and $r$ are related
to the Euler angle parameterization for $SU(2)$ valued fields.

The basic building block of the construction in the Yang-Mills-Higgs case is
the ansatz which describes Hadronic layers in the Skyrme case. To see how
this works, let us consider $U(x)\in SU(2)$ defined as 
\begin{equation}
U=\exp \left( p\,\theta \frac{\mathbf{t}_{3}}{2}\right) \exp \left( H\left(
t,\phi \right) \frac{\mathbf{t}_{2}}{2}\right) \exp \left( q\,r\frac{\mathbf{%
t}_{3}}{2}\right) \,,  \label{ans2}
\end{equation}%
where $p$ and $q$ are non-vanishing integers. The theory of Euler angles for 
$SU(N)$ \cite{euler1,euler2,euler3} dictates the range of $\theta $, $r$ and $%
H $. As it will be discussed in a moment, when $H\left( t,\phi \right) $
satisfies either 
\begin{equation}
H\left( t,\phi =0\right) =0\,,~~~~H\left( t,\phi =\pi \right) =\pi \ ,
\label{ans2.1}
\end{equation}%
or%
\begin{equation}
H\left( t,\phi =0\right) =\pi \ ,\ \ H\left( t,\phi =\pi \right) =0\, .
\label{ans2.01}
\end{equation}%
The Chern-Simons topological charge of the gauge field (to be constructed
in a moment) will be non-zero. To complete the construction of the ansatz
for the gauge potential, we will use the following definitions 
\begin{equation*}
A_{\mu }=\sum_{j=1}^{3}\lambda _{j}\Omega _{\mu }^{j}\mathbf{t}_{j}\,,\qquad
U^{-1}\partial _{\mu }U=\sum_{j=1}^{3}\Omega _{\mu }^{j}\mathbf{t}_{j}\ ,
\label{ans1}
\end{equation*}%
where 
\begin{equation*}
H\left( t,\phi \right) =\arccos \left( G\right) \,,~~G=G\left( t,\phi
\right) \ ,  \label{ans3}
\end{equation*}%
\begin{eqnarray*}
\lambda _{1}\left( t,\phi \right) &=&\lambda _{2}\left( t,\phi \right) =%
\frac{G}{\sqrt{G^{2}+\exp (2\eta )}}\overset{def}{=}\lambda \left( t,\phi
\right) \ ,\ \ \ \lambda _{3}\left( t,\phi \right) =1\ ,\ \eta \in \mathbb{R}%
\ ,  \label{ans4} \\
G\left( t,\phi \right) &=&\exp (3\eta )\frac{F}{\sqrt{1-\exp (4\eta )\cdot
F^{2}}}\,,~~~~F=F\left( t,\phi \right) \ .  \label{ans4.1}
\end{eqnarray*}%
The real parameter $\eta $ will be fixed by requiring that the CS charge is
an integer.

The option in Eq.~\eqref{ans2.01} gives rise to the following boundary
condition for $F\left( t,\phi \right) $, 
\begin{equation}
F\left( t,\phi =0\right) =F_{0}=F\left( t,\phi =\pi \right) \ .
\label{ans2.002}
\end{equation}%
In the latter case the CS charge vanishes. On the other hand, the option in
Eq.~\eqref{ans2.1}, in terms of $F\left( t,\phi \right) $,\ reads 
\begin{equation*}
F\left( t,\phi =0\right) =-\frac{\exp (-2\eta )}{\sqrt{1+\exp (2\eta )}}\
,\qquad F\left( t,\phi =\pi \right) =\frac{\exp (-2\eta )}{\sqrt{1+\exp
(2\eta )}}\ ,  \label{BC}
\end{equation*}%
in order to have a non-zero CS charge. In this case both the CS charge and
the CS density will be non-trivial. Then, $A_{\mu }$ reads%
\begin{align*}
A_{\mu }=& \lambda \left( t,\phi \right) \biggl[\frac{\mathbf{t}_{1}}{2}%
\left\{ -\sin \left( qr\right) dH+p\cos \left( qr\right) \sin \left(
H\right) d\theta \right\} +\frac{\mathbf{t}_{2}}{2}\{\cos \left( qr\right)
dH+p\sin \left( qr\right) \sin \left( H\right) d\theta \}\biggl] + \notag \\
& +~\frac{\mathbf{t}_{3}}{2}\left[ qdr+p\cos (H)d\theta \right] \,,
\label{ansgauge}
\end{align*}%
\begin{equation*}
dH=\frac{\partial H}{\partial t}dt+\frac{\partial H}{\partial \phi }d\phi \ .
\end{equation*}

With the above, the complete set of $(3+1)$-dimensional Yang-Mills field
equations reduce to 
\begin{equation*}
\square F\,\equiv \,\left( \frac{\partial ^{2}}{\partial t^{2}}-\frac{1}{%
L_{\phi }^{2}}\frac{\partial ^{2}}{\partial \phi ^{2}}\right) F\,=\,0\,,
\end{equation*}%
which corresponds to the field equation of a free massless scalar field in
two dimensions.

\subsection{Energy-momentum tensor and topological charge}


With the above ansatz the topological density, the on-shell Lagrangian and
the energy-momentum tensor read 
\begin{equation*}
\rho _{\text{CS}}=\frac{pq\exp (3\eta )}{16\pi ^{2}\left( 1-\exp (4\eta
)F^{2}\right) ^{3/2}}\frac{\partial F}{\partial \phi }\,,  
\end{equation*}%
\begin{align*}
T_{tt}=& \frac{p^{2}}{e^{2}L_{\theta }^{2}}\exp (5\eta )\cosh \left( \eta
\right) \left[ \left( \frac{\partial F}{\partial t}\right) ^{2}+\frac{1}{%
L_{\phi }^{2}}\left( \frac{\partial F}{\partial \phi }\right) ^{2}\right] \ ,\\
L_{\text{on-shell}}=& \frac{p^{2}}{e^{2}L_{\theta }^{2}}\exp (5\eta )\cosh
\left( \eta \right) \left[ \left( \frac{\partial F}{\partial t}\right) ^{2}-%
\frac{1}{L_{\phi }^{2}}\left( \frac{\partial F}{\partial \phi }\right) ^{2}%
\right] \ .  
\end{align*}%
The full energy-momentum tensor is 
\begin{equation*}
T_{\mu \nu }=\left[ 
\begin{array}{cccc}
T_{tt} & 0 & 0 & P_{\phi } \\ 
0 & T_{rr} & 0 & 0 \\ 
0 & 0 & T_{\theta \theta } & 0 \\ 
P_{\phi } & 0 & 0 & T_{\phi \phi }%
\end{array}%
\right] \,,
\end{equation*}%
where%
\begin{align*}
T_{rr}=& \frac{p^{2}L_{r}^{2}}{e^{2}L_{\theta }^{2}}\exp \left( 5\eta
\right) \cosh \left( \eta \right) \left[ \left( \frac{\partial F}{\partial t}%
\right) ^{2}-\frac{1}{L_{\phi }^{2}}\left( \frac{\partial F}{\partial \phi }%
\right) ^{2}\right] =-\frac{L_{r}^{2}}{L_{\theta }^{2}}T_{\theta \theta }\ ,\\
T_{\phi \phi }=& \frac{p^{2}L_{\phi }^{2}}{e^{2}L_{\theta }^{2}}\exp \left(
5\eta \right) \cosh \left( \eta \right) \left[ \left( \frac{\partial F}{%
\partial t}\right) ^{2}+\frac{1}{L_{\phi }^{2}}\left( \frac{\partial F}{%
\partial \phi }\right) ^{2}\right] \,,  
\end{align*}%
\begin{equation*}
T_{t\phi }=P_{\phi }=\frac{2p^{2}\exp \left( 5\eta \right) \cosh \left( \eta
\right) }{e^{2}L_{\theta }^{2}}\frac{\partial F}{\partial t}\frac{\partial F%
}{\partial \phi }\,.  \label{tmunu3}
\end{equation*}%
Of course, the energy-momentum tensor is traceless; $g^{\mu \nu }T_{\mu \nu }=0$%
. Just as in the cases of Hadronic layers and tubes described in previous sections, the effective two-dimensional energy-momentum tensor
restricted to the $t$ and $\phi $\ directions are still traceless: 
\begin{equation*}
T_{ab}=\left( 
\begin{array}{cc}
T_{tt} & P_{\phi } \\ 
P_{\phi } & T_{\phi \phi }%
\end{array}%
\right) \ ,\ \qquad a,b=t,\phi \,.
\end{equation*}
On the other hand, the topological $Q_{CS}$ charge is 
\begin{align*}
Q_{CS} &= \left. \frac{pq \exp{(3\eta)}}{2}\left[ \frac{F}{\sqrt{1 - \exp{(4
\eta)}F^{2}}} \right] \right|^{F(t,\pi)}_{F(t,0)}.
\end{align*}
We will consider the boundary conditions of \eqref{ans2.002} for $F(t,\phi)$%
, because in Eq. \eqref{ans2.002} it was concluded that for values $F(t,0) =
F(t, \pi)$ the topological charge is canceled. Now, we will introduce an
auxiliary function for the $B_{CS}$ to be expressed in terms of an integer. 
\begin{align*}
\Omega(\eta,a,b) \equiv \frac{\exp(3\eta)}{2}\left[ \frac{a}{\sqrt{1-\exp{%
(4\eta)}a^2 }} - \frac{b}{\sqrt{1-\exp{(4\eta)}b^2 }} \right].
\end{align*}
Then, the topological charge is the following 
\begin{align*}
Q_{CS} = pq \cdot \Omega(\eta, a= F(t,\pi), b= F(t,0)).
\end{align*}
Now if we consider the boundary conditions \eqref{ans2.002} for $F(t,\phi)$
in the auxiliary function $\Omega$ we obtain the following reduced
expression for the topological charge 
\begin{align*}
Q_{CS} &= pq.  \label{aux2}
\end{align*}
Thus $pq$ must be integers because the $Q_{CS}$ must be an integer.

\subsection{Yang-Mills-Higgs theory}

The previous results can be extended to the Yang-Mills-Higgs theory in $%
(3+1) $-dimensions (with the Higgs field in the adjoint representation)
whose action reads 
\begin{equation*}
I[A,\varphi ]=\int d^{4}x\sqrt{-g}\,\biggl(\frac{1}{2e^{2}}\text{Tr}(F_{\mu
\nu }F^{\mu \nu })+\frac{1}{4}\text{Tr}(D_{\mu }\varphi D^{\mu }\varphi )%
\biggl)\,.  \label{IH2a}
\end{equation*}%
The field equations and the energy-momentum tensor are 
\begin{gather*}
\nabla _{\nu }F^{\mu \nu }+[A_{\nu },F^{\mu \nu }]+\frac{e^{2}}{4}[\varphi
,D^{\mu }\varphi ]\ =\ 0\ ,  \label{Eq2a} \\
D_{\mu }D^{\mu }\varphi \ =\ 0\ ,  \label{EqHa}
\end{gather*}
\begin{equation*}
T_{\mu \nu }=-\frac{2}{e^{2}}\text{Tr}\biggl(F_{\mu \alpha }{F_{\nu }}%
^{\alpha }-\frac{1}{4}g_{\mu \nu }F_{\alpha \beta }F^{\alpha \beta }\biggl)-%
\frac{1}{2}\text{Tr}\biggl(D_{\mu }\varphi D_{\nu }\varphi -\frac{1}{2}%
g_{\mu \nu }D_{\alpha }\varphi D^{\alpha }\varphi \biggl)\ .
\end{equation*}%
Also in the present case, the main building block is the ansatz for the
gauge potential ``inspired" by the expression for the Hadronic layers of the
Skyrme model (described in the previous subsection). On the other hand, the
Higgs field can be chosen as 
\begin{equation*}
\varphi =\sum_{j=1}^{3}f_{j}(r)h^{j}(t,\phi )\mathbf{t}_{j}\ ,
\label{varphi}
\end{equation*}%
\begin{gather*}
h_{1}(t,\phi )=\frac{a}{b}h(t,\phi )\ ,\quad h_{3}(t,\phi )=a\cot \left(
H(t,\phi )\right) \frac{h(t,\phi )}{\lambda (t,\phi )}\ ,\quad \lambda
_{3}=1\ ,  \label{hs} \\
f_{1}(r)=b\cos (qr)f_{3}(r)\ ,\quad f_{2}(r)=a\sin (qr)f_{3}(r)\ ,\quad
f_{3}(r)=f_{0}\,r\ ,  \label{fs}
\end{gather*}
\begin{equation*}
h_{2}(t,\phi ):=h(t,\phi )\ ,\qquad \lambda _{1}(t,\phi )=\lambda
_{2}(t,\phi ):=\lambda (t,\phi )\ ,
\end{equation*}%
where $a$, $b$, and $f_{0}$ are integration constants.

In this way, the Yang-Mills-Higgs equations read 
\begin{equation*}
\Box H=\left( \frac{\partial ^{2}}{\partial t^{2}}-\frac{1}{L_{\phi }^{2}}%
\frac{\partial ^{2}}{\partial \phi ^{2}}\right) H=0\,,\qquad \Box h=\left( 
\frac{\partial ^{2}}{\partial t^{2}}-\frac{1}{L_{\phi }^{2}}\frac{\partial
^{2}}{\partial \phi ^{2}}\right) h=0\,,\qquad \Box \lambda =\left( \frac{%
\partial ^{2}}{\partial t^{2}}-\frac{1}{L_{\phi }^{2}}\frac{\partial ^{2}}{%
\partial \phi ^{2}}\right) \lambda =0\,,
\end{equation*}%
together with 
\begin{eqnarray*}
\left( \frac{\partial H}{\partial t}\right) ^{2}-\frac{1}{L_{\phi }^{2}}%
\left( \frac{\partial H}{\partial \phi }\right) ^{2} &=&\left( \frac{%
\partial H}{\partial t}-\frac{1}{L_{\phi }}\frac{\partial H}{\partial \phi }%
\right) \left( \frac{\partial H}{\partial t}+\frac{1}{L_{\phi }}\frac{%
\partial H}{\partial \phi }\right) =0\,, \\
\left( \frac{\partial h}{\partial t}\right) ^{2}-\frac{1}{L_{\phi }^{2}}%
\left( \frac{\partial h}{\partial \phi }\right) ^{2} &=&\left( \frac{%
\partial h}{\partial t}-\frac{1}{L_{\phi }}\frac{\partial h}{\partial \phi }%
\right) \left( \frac{\partial h}{\partial t}+\frac{1}{L_{\phi }}\frac{%
\partial h}{\partial \phi }\right) =0\,, \\
\left( \frac{\partial \lambda }{\partial t}\right) ^{2}-\frac{1}{L_{\phi
}^{2}}\left( \frac{\partial \lambda }{\partial \phi }\right) ^{2} &=&\left( 
\frac{\partial \lambda }{\partial t}-\frac{1}{L_{\phi }}\frac{\partial
\lambda }{\partial \phi }\right) \left( \frac{\partial \lambda }{\partial t}+%
\frac{1}{L_{\phi }}\frac{\partial \lambda }{\partial \phi }\right) =0\,,
\end{eqnarray*}
\begin{equation*}
~\lambda =\pm \frac{\cos (H)}{\sqrt{\exp (2\lambda _{0})+\cos ^{2}(H)}}\ .
\label{lambdasol}
\end{equation*}

Summarizing, with the ansatz presented above, the complete set of field
equations of the Yang-Mills-Higgs theory has been reduced to the field
equations of three chiral massless scalar fields in $(1+1)$-dimensions plus
a non-linear constraint between the two of them. Consequently, the
techniques that have been developed to analyze the ``dressing" of Hadronic
tubes and layers with chiral modes (allowing, in principle, the computations of relevant
transport coefficients) also, work in the Yang-Mills-Higgs case.

\section{Conclusion}

This review describes a proper analytic framework to construct inhomogeneous Baryonic condensates in the gauged Skyrme Maxwell theory. This approach cannot only produce exact sol

utions with high Baryonic charges but also gives considerable physical insights into the nature of the configurations (such as the fact that these Hadronic layers and tubes constructed in the previous sections are natural sources of force free plasmas). Another characteristic of the present technique is that it discloses the appearance of chiral conformal degrees of freedom which describes modulations of the condensates. Finally, we have discussed that a similar strategy also works in the case of Yang-Mills-Higgs theory in $(3+1)$ dimensions where the insights from the gauged Skyrme model help to construct novel analytic solutions (which can also be dressed with chiral conformal
degrees of freedom).

\subsection*{Acknowledgements}
F. C. has been funded by Fondecyt Grants 1200022. S. R. has been funded by ANID-Beca de Magíster Nacional 2022-22221100, FONDECYT Grant 221504, and 1200022. The Centro de Estudios Cient%
\'{\i}ficos (CECs) is funded by the Chilean Government through the Centers
of Excellence Base Financing Program of ANID.

\newpage %

\end{document}